# The automatic detection of the information operations event basis

D.V. Lande, C.V. Prishchepa

Institute for Information Recording of NAS of Ukraine

The methodology of automatic detection of the event basis of information operations, reflected in thematic information flows, is described. The presented methodology is based on the technologies for identifying information operations, the formation of the terminological basis of the subject area, the application of cluster analysis with cluster centroids, determined by analyzing the terminology of the information flow. The clusters formed in this way reflect the main events occurring during the information operations and reveal the technique for their implementation.

### Introduction

At present, the Internet space becomes a battlefield, on which there are numerous information wars, individual information operations. Information operations are defined as "actions aimed at influencing information and information systems of the enemy and protecting their own information and information systems" [1]. Information operations are components and support for more general processes. At the same time, the arena of information operations is the information space, which, on the one hand, is the place of information battles, and on the other hand, the environment for displaying real combat operations [2]. In this case, information operations in practice are supported by numerous events, processes, actions (under the events within the framework of this work we will understand a significant incident, phenomenon or manifestation of other activity as a fact of public or private life). Analysis of the reflection of events in the Internet space makes it possible to identify the participants in information confrontations, methods of information impact, to uncover the technique of implementing information operations.

### Objective

The purpose of this work is to create and justify a methodology for identifying the event basis of information operations. The implementation of this methodology will let us determine the time frame of the information operation, identify the main events that accompany the information operation and see the techniques of information impact.

When studying information operations it is necessary to determine objective criteria, and as one such, one can consider the dynamics of the distribution of information plots in the corresponding fragment of the information space. Numerous scientific works [3], [4], [5], [6] have been devoted to the investigation of the dynamics of information flows, it is shown that in typical situations the dynamics of the distribution of news, information plot is characterized by the nature of a "burst", waves with an obvious period of increasing its influence and smooth decrease. At the same time, the question of determining the event basis for information operations remains open.

**The dynamics curve of the information operation**

It is assumed that a systematic violation of the typical dynamics of some thematic information flows in the open information space may indicate information operations [7]. In the study of information operations, much attention is also paid to the analysis of the dynamics of information flows, in [8] a typical template of the information operation is presented (Figure 1), which allows using available analytical tools, for example, correlation analysis [9].

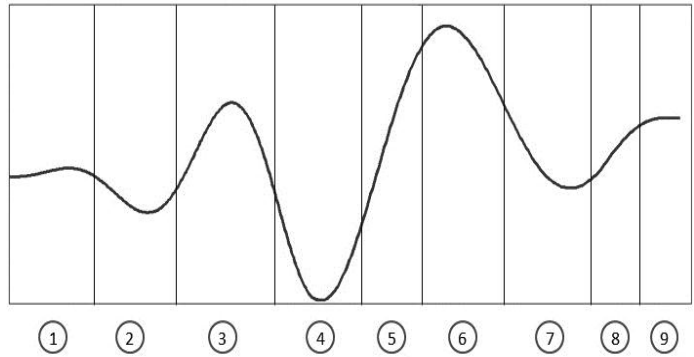

Fig. 1 - Generalized diagram, corresponding to all stages of the life cycle of information operations: 1 - background; 2 - calm; 3 - "art preparation"; 4 - calm; 5 - Attack / Growth Trigger; 6 - peak of high expectations; 7 - loss of illusions; 8 - public awareness; 9 - productivity / background

**Using the content monitoring system**

To get the dynamics of the thematic flow on a certain topic, you can use content monitoring systems. As the system of content monitoring, the authors selected the InfoStream system, which currently covers 10,000 sources of information in Russian, Ukrainian and English. The database of the system receives more than 100 thousand documents daily. The InfoStream system provides a search, as well as viewing the list and full texts of relevant documents.

In the example shown in Fig. 2 a fragment of the system interface through which the request for the referendum on the UK exit from the European Union (abbreviated Brexit from the combination of the words Britain - Britain and English Exit - exit) processed during 2016 (the period of the study June-July 2016) is shown. As a result, a thematic information array was created, covering 43697 documents. In Fig. 3 a graph of the dynamics of this information flow, as well as the result of its smoothing with a window in 7 days is shown.

![Fragment of content monitoring system interface showing Brexit-related news articles]

Fig. 2 - Fragment of the interface of the content monitoring system

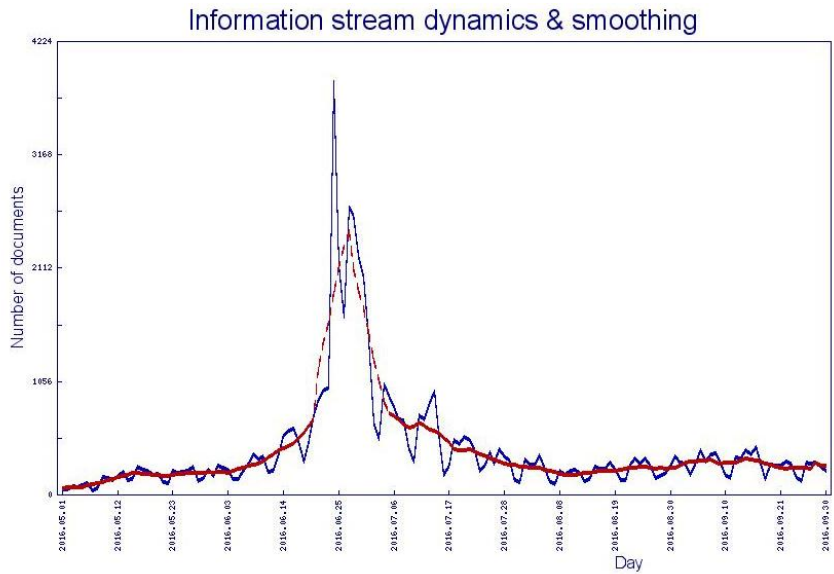

Fig. 3 - Dynamics of the thematic information flow (Brexit)

**Correlation with the information operation pattern**

To determine the degree of "proximity" of the dynamics of the thematic information flow to the information operation, the method proposed in [9] was used. The idea is to compare parts of a series of information flow dynamics with a certain pattern on different scales. For this, the correlation between the time series part and some template (the scaled part of the graph shown in Figure 1) is calculated:

$$C(l,k) = \frac{\sum_{i=1}^{k}(x_{l+i} - \bar{x})(p_i - \bar{p})}{\sqrt{\sum_{i=1}^{k}(x_{l+i} - \bar{x})^2 \sum_{i=1}^{k}(p_i - \bar{p})^2}},$$

The resulting coefficient $C(l,k)$ depends on the values $x_{l+1},...,x_{l+k}$. That is, the parameter $k$ corresponds to the pattern shift, and the parameter $k$ corresponds to the number of points in the pattern and in the considered segment of the row. The parameter $k$ in this case is analogous to the scale. In this case, for calculation $C(l,k)$ the $k$ points of the series and the length pattern are used. When visualizing, the most light colors correspond to the highest values. Fig. 4 shows a Correlogram of the time series corresponding to the dynamics of the information flow on the Brexit topic and the template shown in Fig. 1.

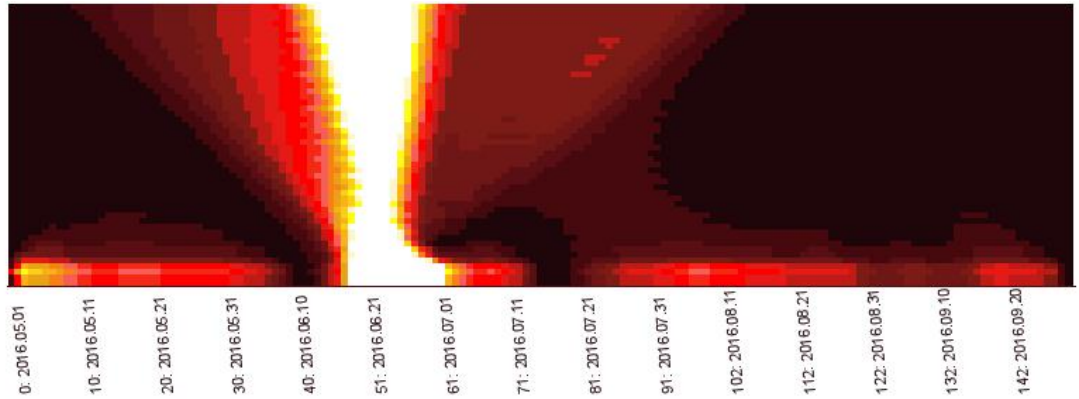

Fig. 4. Correlation coefficients of the thematic information flow

with information operation patterns

The analysis of the given correlogram makes it possible to narrow the time frame in the study of the thematic information flow.

**Terminological basis of the subject domain**

To determine the event basis of information operations, reference words are extracted from the information flow documents, for which several algorithms can be used [10]. The authors, in particular, used the TF-IDF algorithm implemented in the InfoStream system. Then, the significant information words for the information flow were automatically compared with the dictionary of words developed by the authors, which describe the events. As a result of the analysis of the information flow on Brexit, the following words were identified: PROTEST, REFERENDUM, PETITION, SIGNATURES, DEMONSTRATION, TERRORIST ACT ...

The basic query for the formation of the information flow according to the information set to the InfoStream system was supplemented with these words, which allowed to limit the number of documents, leaving only those that describe specific events. This also made it possible to determine the list of information sources that publish the event information and build a network of their interrelations (Figure 5).

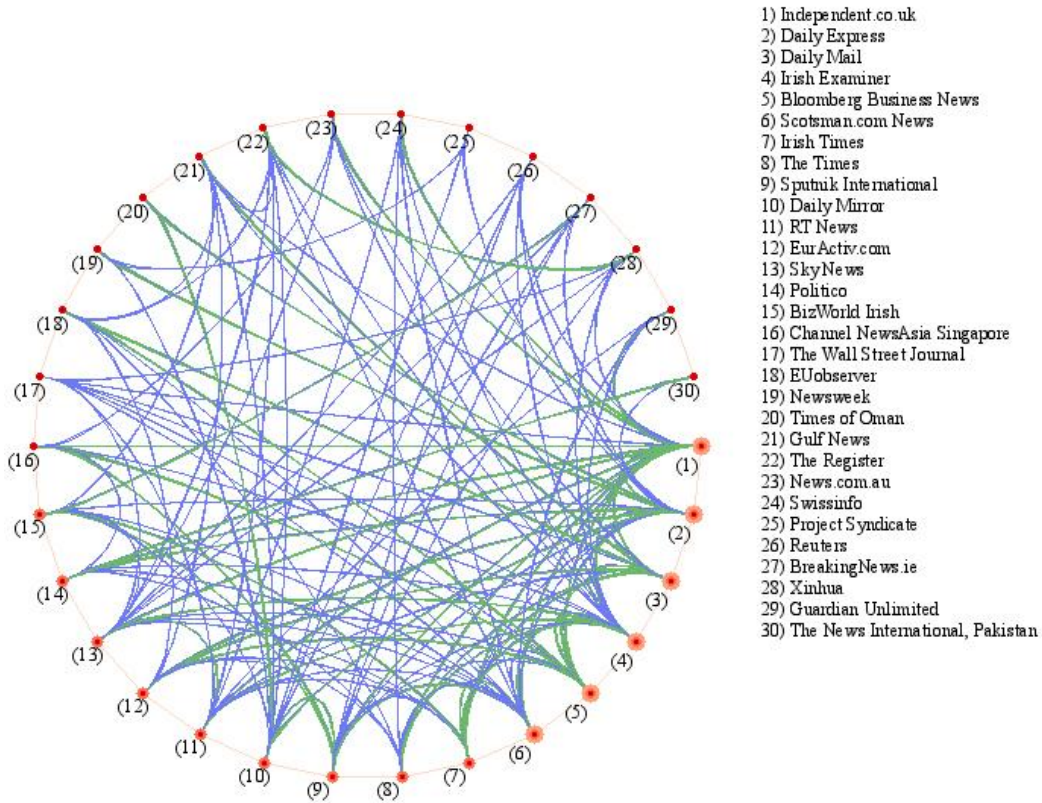

Fig. 5. A fragment of the graph of horizontal visibility, reflecting the links of sources of information on the given topic (Brexit)

**Clustering of the domain**

In order to identify clusters of the main events, the received information flow is analyzed using the k-means method of document packing into a fixed number of clusters $\{d^{(1)},...,d^{(N)}\}$, which consists in the following: the $k$ documents that are defined as centroids (the most typical representatives of clusters) are selected in some ways. That means that each cluster $C_j$ ($j = 1, ..., k$) is represented by the corresponding centroid. The closeness of documents to the centroid can be determined in different ways, with the help of some measure $Sim(d, C_j)$.

After that, the document is attributed to the cluster, the value $Sim(d, C_j)$ for which is the largest. Then, for each of the new clusters, the centroid $C_j$ ($j = 1, ..., k$) is newly defined, for example, as a set of the most significant (by some selected criterion) words from the documents included in this cluster.

After that, the process of filling the clusters, then the calculation of new centroids, etc., continues until the process of forming clusters stabilizes.

It is known that the k-means algorithm maximizes the quality function of clustering $Q$:

$$Q(C_1,...,C_k) = \sum_{j=1}^{k} \sum_{d \in C_j} Sim(d, C_j).$$

This method has a high speed (complexity of the order $O(k, N)$, where $k$ is the required number of clusters, $N$ is the number of documents). At the same time, when using this method an open question always remains the choice of starting documents for clustering - centroids. The authors proposed to use as the centroids earlier determined key words, which express the essence of the events. The proposed approach allowed simple implementation of clustering within the capabilities of the InfoStream system (Fig. 6).

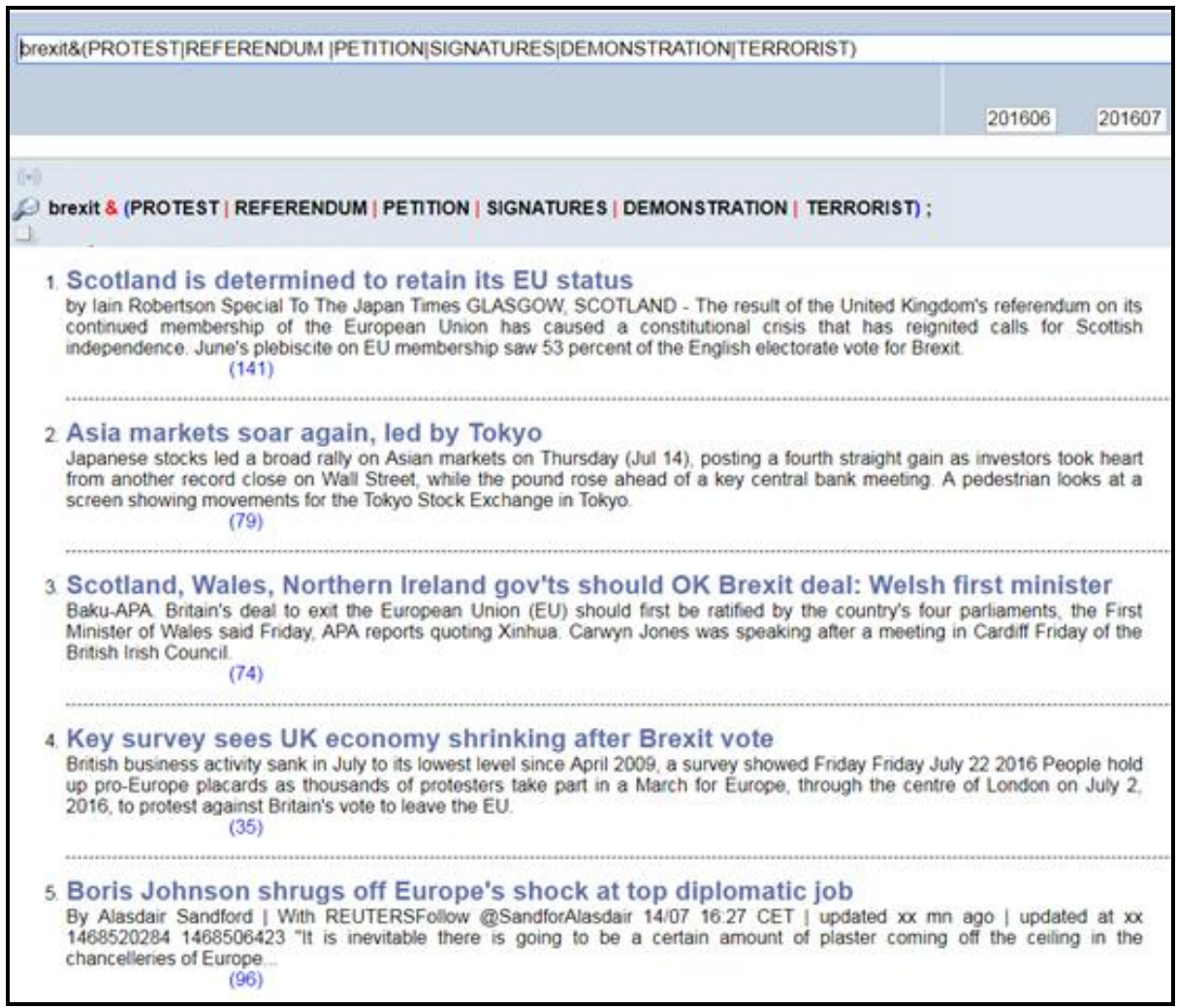

Fig. 6. The main plots (clusters) of the thematic information flow

**Conclusions**

Thus, the proposed method for identifying the event basis of information operations covers the following stages of the research, the result of each of them can be considered as an independent information product:

1. Formation of a thematic information flow on the topic

2. Investigation of the dynamics of the received flow, identification of signs of an information operation

3. Narrowing the time frame of the information flow, obtaining a more representative sample of documents.

4. Definition of the terminological basis for the description of events within the scope of the subject area under study. Identify sources of information about events (if necessary).

5. Clustering, identifying the main events that accompany the information operation

The clusters of documents determined as a result of the proposed methodology correspond to the main events that accompany the information operation. Further meaningful analysis of these clusters makes it possible to identify the main characters of the information operation, to disclose the technique of its implementation.

List of references